\documentclass[11pt,twoside]{article}  
\usepackage{adassconf}

\begin{document}   

%
%

\paperID{O8.1}

%
%
%
%

\title{Using the Parallel Virtual Machine for Everyday Analysis}

%
%
%

\author{M. S. Noble, J. C. Houck, J. E. Davis, A. Young, M. Nowak}
\affil{Kavli Institute for Astrophysics and Space Research, Massachusetts Institute of Technology}

%
%

\contact{Michael S. Noble}
\email{mnoble@space.mit.edu}

%
%
%
%
%

\paindex{Noble, M.}
\aindex{Houck, J.}
\aindex{Davis, J.}
\aindex{Young, A.}
\aindex{Nowak, M.}

%
%

\authormark{Noble et al.}

%
%



\begin{abstract}          
A review of the literature reveals that while parallel computing is
sometimes employed by astronomers for custom, large-scale calculations,
no package fosters the routine application of parallel methods to
standard problems in astronomical data analysis.
This paper describes our attempt to close that gap by wrapping the
Parallel Virtual Machine (PVM) as a scriptable S-Lang module.  Using
PVM within ISIS, the Interactive Spectral Interpretation System, we've
distributed a number of representive calculations over a network of
25+ CPUs to achieve dramatic reductions in execution times.
We discuss how the approach applies to a wide class of modeling problems,
outline our efforts to make it more transparent for common use, and note
its growing importance in the context of the large, multi-wavelength
datasets used in modern analysis.
\end{abstract}

%
%

\section{Introduction} 

Parallel computing is not a new discipline, so it is surprising that
few astronomers resort to parallelism when solving standard problems
in data analysis.  To quantify this assertion relative to the X-ray
community, in late summer of 2005 we conducted several full text searches
of the NASA ADS digital library (Kurtz et al 1993), as follows:\\

\begin{tabular}{cc}
Keywords & Number of Hits \\
\hline
parallel AND pvm & 38  \\
message AND passing AND mpi & 21 \\
xspec & 832 \\
xspec AND parallel AND pvm & 0 \\
xspec AND message AND passing AND mpi & 0 \\\\
\end{tabular}

\noindent
Extra keywords were included with PVM and MPI so as to
cull false matches (e.g. with the Max Planck Institute).  The keyword
{\tt xspec} refers to the software program of the same name
(Arnaud 1996), which is generally regarded as the most
widely used application for modeling X-ray spectra.  Queries in ADS
on other modeling tools, or with other search engines such as Google,
all yield similar
trends: astronomers and astrophysicists do employ parallel computing,
but mainly for highly customized, large-scale problems in simulation,
image processing, or data reduction.  Virtually no one is using parallelism
for fitting models within established software systems, especially in the
interactive context, even though a majority of papers published in
observational astronomy result from exactly this form of analysis.

\section{ISIS, S-Lang, PVM, and SLIRP}

To exploit this opportunity we've extended ISIS, the Interactive Spectral
Interpretation System (Houck 2002), with a dynamically importable module
that provides scriptable access to the Parallel Virtual Machine (Geist
et al 1994). PVM was selected (e.g. over MPI) for its robust fault
tolerance in a networked environment.  ISIS, in brief, was originally
conceived as a tool for
analyzing Chandra grating spectra, but quickly grew into a
general-purpose analysis system.  It
provides a superset of the XSpec models and, by embedding the S-Lang
interpreter, a powerful scripting environment complete with fast
array-based mathematical capabilities rivaling commercial packages such
as MatLab or IDL.
Custom user models may be loaded into ISIS as either scripts
\footnote{Usually in S-Lang, but Python may also be used
by simply importing the PySL module.} or compiled code, without any recompilation of ISIS
itself; because of the fast array manipulation native to S-Lang, scripted
models suffer no needless performance penalties, while the SLIRP code
generator (Noble 2003) can render the use of compiled C, C++, and FORTRAN
models a nearly instantaneous, turnkey process.

\section{Parallel Modeling}
Using the PVM module we've parallelized a number of the numerical modeling
tasks in which astronomers engage daily, and summarize them here as a series
of case studies. Many of the scientific results stemming from these efforts
are already appearing elsewhere in the literature.

\subsection{Kerr Disk Line}
Relativistic Kerr disk models are computationally expensive.  Historically,
implementors have opted to use precomputed tables to gain speed at the
expense of limiting flexibility in searching parameter space.  However, by
recognizing that contributions from individual radii may be computed
independently we've parallelized the model to avoid this tradeoff.  To
gauge the performance benefits 
\footnote{A more complete and rigorous analysis will be presented in a
future journal paper.}
we tested the sequential execution of a single model evaluation, using
a small, faked test dataset, on our fastest CPU (a 2Ghz AMD Opteron),
yielding a median runtime of 33.86 seconds.  Farming the same computation
out to 14 CPUs on our network reduced the median runtime to 8.16s, yielding
a speedup of 4.15.  While 30\% efficiency seems unimpressive at first
glance, this result actually represents 67\% of the peak speedup of 6.16
predicted by Amdahl's Law
(5.5 of the 33.86 seconds runtime on 1 CPU was not parallelizable in the
 current implementation), on CPUs of mixed speeds and during normal
working hours. Reducing the model evaluation time to 8 seconds brings it
into the realm of interactive use,
with the result that fits requiring 3-4 hours to converge (on "real" datasets
such as the long XMM-Newton observation of MCG--6-30-15 by Fabian) may
now be done in less than 1 hour.  The model evaluation is initiated in ISIS
through the S-Lang hook function
\begin{verbatim}
     public define pkerr_fit (lo, hi, par)
     {
         variable klo, khi;
         (klo, khi) = _A(lo, hi);	% convert angstroms to KeV
         return par[0] * reverse ( master (klo, khi, par));
     }
\end{verbatim}
where {\tt lo} and {\tt hi} are arrays (of roughly 800 elements) representing
the left and right edges of each bin within the model grid, and {\tt par} is
a 10 element array of the Kerr model parameters.  Use of the PVM module is
hidden within the {\tt master} call (which partitions the disk radii
computation into slave tasks), allowing ISIS to remain unaware that the
model has even been parallelized.  This is an important point: {\itshape
parallel models are installed and later invoked using precisely the same
mechanisms employed for sequential models.}
\footnote{This also makes it easy for ISIS to employ an MPI module for
parallelism, if desired.}
For each task the slaves invoke a FORTRAN {\tt kerr} model implementation,
by Laura Breneman at the University of Maryland, wrapped by SLIRP as follows:
\begin{verbatim}
     linux%   slirp -make kerr.f
     Starter make file generated to kerr.mf
     linux%   make -f kerr.mf
\end{verbatim}
%
\subsection{Confidence Contours and Error Bars}
Error analysis is ripe for exploitation with parallel methods.  In
the 1D case, an independent search of $\chi^2$ space may be made for each
of the {\tt I} model parameters, using {\tt N=I}\  slaves, with each
treating one parameter as thawed and {\tt I-1} as fixed.  Note that
superlinear speedups are possible here, since a slave finding a lower
$\chi^2$ value can immediately
terminate its {\tt N-1} brethren and restart them with updated parameters
values.  Parallelism in the 2D case is achieved
by a straightforward partition of the parameter value grid into {\tt J}
independently-evaluated rectangles, where {\tt J} $>>$ {\tt N} (again,
the number
of slaves) is typical on our cluster.  Our group and collaborators have
already published several results utilizing this technique. For example,
Allen et al 2004 describes joint X-ray, radio, and $\gamma$-ray fits of
SN1006, containing a synchrotron radiation component modeled as
\begin{displaymath}
\frac{dn}{dkdt} = 
\frac{\sqrt{3}e^{3} B}{hmc^{2}k} \int{dpN(p)R \left (\frac{k}{k_0\gamma^2} \right )}
\end{displaymath}

\noindent
The physics of this integral is not important here; what matters is that
the cost of evaluating it over a 2D grid is prohibitive (even though
symmetry and precomputed tables have reduced the integral from 3D to 1D),
since it must be computed once per spectral bin, hundreds of times
per model evaluation, and potentially millions of times per confidence grid.
A 170x150 contour grid (of electron spectrum exponential cutoff energy
versus magnetic field strength) required 10 days to compute on 20-30
CPUs (the fault tolerance of PVM is critical here),
and would scale linearly to a 6-10 month job on a single workstation.

\subsection{Temperature Mapping}
Temperature mapping is another problem that is straightforward to
parallelize and for which we have already published results.  For
instance, Wise \& Houck
2004 provides a map of heating in the intracluster medium of Perseus,
computed from 10,000 spectral extractions and fits on 20+ CPUs
in just several hours.

\section{Going Forward}
It is important to note that in the two previous studies {\itshape 
the models themselves were not parallelized}, so the usual entry barrier of
converting serial codes to parallel does not apply.  One consequence
is that the community should no longer feel compelled to compute error
analyses or temperature maps serially.  Another consequence is that
the independence between partitions of the data and the computation
being performed, which makes the use of sequential models possible in
the parallel context, also lurks within other areas of the modeling
problem.  In principle it should be possible to evaluate an arbitrary
sequential model in parallel by partitioning the model grid over which
it's evaluated, or by evaluating over
each dataset independently (when multiple datasets are fit), or in
certain cases even by evaluating non-tied components in parallel. We
are implementing these techniques with an eye towards rendering their
use as transparent as possible for the non-expert.
With simple models or small datasets these measures may be not be necessary,
but the days of simple models and small datasets are numbered.  Reduced
datasets have already hit the gigabyte scale, and multi-wavelength analysis
such as we describe above is fast becoming the norm.  These trends will
only accelerate as newer instruments are deployed and the Virtual Observatory
is more widely utilized, motivating scientists to tackle more ambitious
analysis problems that may have been shunned in the past due to their
computational expense.

\acknowledgments
This work was supported by NASA through the AISRP grant NNG05GC23G
and Smithsonian Astrophysical Observatory contract SV3-73016 for the
Chandra X-Ray Center.


\scriptsize

\begin{references}
\reference Allen, G.~E., Houck, J.~C., \& Sturner, S.~J.\ 2004, Advances in
Space Research, 33, 440
\reference Kurtz, M.J., Karakashian, T., Grant, C.S., Eichhorn, G., Murray, S.S., Watson, J.M., Ossorio, P.G., \& Stoner, J.L.\ 1993 \adassii
\reference Arnaud, K.A.\ 1996 \adassv
\reference A. Geist,  A. Beguelin,  J. Dongarra, W. Jiang, R. Manchek, \&
	V. Sunderam 1994, PVM: Parallel Virtual Machine, A User's Guide and Tutorial for Networked Parallel Computing
\reference Houck, J.~C.\ 2002, ISIS: The Interactive Spectral Interpretation System, High Resolution X-ray Spectroscopy with XMM-Newton and Chandra
\reference Noble, M.~S.\ 2003, http://space.mit.edu/cxc/software/slang/modules/slirp
\reference Wise, M., \& Houck, J.\ 2004, 35th COSPAR Scientific Assembly, 3997
\end{references}
\end{document}